\documentclass[preprint,12pt]{elsarticle}

\usepackage{amssymb}
\usepackage{amsmath}
\usepackage[section]{placeins}
\newcounter{bla}

\journal{Computer Physics Communications}

\begin{document}

\begin{frontmatter}

\title{Aligning Active Particles Py Package}

\author{Rüdiger Kürsten\corref{author}}

\cortext[author] {Corresponding author.\\\textit{E-mail address:} kursten@ub.edu}
\address{Departament de Física de la Matèria Condensada, Universitat de Barcelona, Martí i Franquès 1, 08028 Barcelona, Spain}
\address{Universitat de Barcelona Institute of Complex Systems (UBICS), 08028 Barcelona, Spain}
\address{Institut für Physik, Universität Greifswald, Felix-Hausdorff-Str. 6, 17489 Greifswald, Germany}

\begin{abstract}
The package performs molecular-dynamics-like agent-based simulations for models of aligning self-propelled particles in two dimensions such as e.g. the seminal Vicsek model or variants of it.
In one class of the covered models, the microscopic dynamics is determined by certain time discrete interaction rules.
Thus, it is no Hamiltonian dynamics and quantities such as energy are not defined.
In the other class of considered models (that are generally believed to behave qualitatively the same) Brownian dynamics is considered.
However, also there, the forces are not derived from a Hamiltonian.
Furthermore, in most cases, the forces depend on the state of all particles and can not be decomposed into a sum of forces that only depend on the states of pairs of particles.
Due to the above specified features of the microscopic dynamics of such models, they are not implemented in major molecular dynamics simulation frameworks to the best of the authors knowledge.
Models that are covered by this package have been studied with agent-based simulations by dozens of papers.
However, no simulation framework of such models seems to be openly available.
The program is provided as a Python package. The simulation code is written in C. In the current version, parallelization is not implemented.

\end{abstract}


\end{frontmatter}

{\bf PROGRAM SUMMARY}

\begin{small}
\noindent
{\em Program Title: aligning active particles py package}                                          \\
{\em Developer's repository link:} https://github.com/kuersten/aappp \\
{\em Licensing provisions:} MIT  \\
{\em Programming language:}      C / Python                             \\
{\em Nature of problem:} Perform molecular-dynamics-like agent-based simulations of models for aligning active particles with interaction rules that are not following Hamiltonian dynamics and that are not restricted to pair-interactions.\\
{\em Solution method:} Uses cell lists to find interacting particles.\\
{\em Additional comments including restrictions and unusual features:} Does not run in parallel. Allows the usage of reflecting boundary conditions.\\
\end{small}

   \section{Introduction}
Active particles are characterized by the transformation of energy into directed motion as well as the dissipation of energy towards their surrounding.
There are engineered active particles such as active colloids or robots, as well as biological active particles from the scales of microtubuli driven by molecular motors towards macroscopic animals, see e.g. \cite{Ramaswamy10, MJRLPRS13, Ramaswamy17, BGHP20, Chate20, SSBMV22, ACJ22} for reviews.

Due to the interplay of constant supply and dissipation of energy, active particles are driven far from thermal equilibrium.
For the description of collective phenomena of active particles, nonequilibrium theories are necessary \cite{Ramaswamy10, MJRLPRS13, Ramaswamy17}.
It is suitable to accompany and test such theories by computer simulations.
However, in order to compare to coarse grained theories one requires to simulate a large number of particles.
Developing and simulating realistic models of a large number of active particles seems to be challenging.

Alternatively, simplified prototype models of self-propelled particles have been introduced, most prominently the famous Vicsek model \cite{VCBCS95}, see also \cite{BGHP20, Chate20} for a review of similar models.
Such simplified models can easily be simulated on much larger scales than realistic models while they keep some key aspects of the far from equilibrium collective motion. 

The aappp simulation package presented here, in the following called the \textit{program}, is able to simulate the two dimensional Vicsek model as well as many similar models.
Models that are covered by the program have been studied by dozens to hundreds of papers, cf. the references of \cite{BGHP20, Chate20}.
However, to the best of the authors knowledge, no simulation framework that can handle those models is openly available.
The models under consideration are not following Hamiltonian dynamics and furthermore they involve $N$-particle interactions of a special type.
For those reasons they are not implemented in well-known molecular dynamics simulation packages.

The source code of the program is available at \cite{aappp}.

This paper is organized as follows.
In Sec.~\ref{sec:models} we define the models that can be handled by the program.
In Sec.~\ref{sec:implementation} we shortly describe the implementation as well as the observables that can be measured automatically on the fly. 
In Sec.~\ref{sec:usage} we give very few remarks on the basic usage of the program.
More instructions can be found in the documentation at \cite{aappp}.
In Sec.~\ref{sec:examples} we display simulation results that have been obtained by the program and compare them to results of previous studies.
Furthermore we show the scaling of the run time.
In Sec.~\ref{sec:versions} we shortly discuss the difference between different versions of the program.
We conclude with a short summary in Sec.~\ref{sec:summary}.

\section{Models\label{sec:models}}
   We consider $N$ point particles in two dimensions.
   The state of each particle $i\in \{1, \cdots, N\}$ is characterized by $x_i\in [0, L_x]$, $y_i\in [0, L_y]$ and $\theta_i \in [-\pi, \pi]$, where $(x_i, y_i)$ gives the position within a two-dimensional simulation box of size $L_x\times L_y$ and the angle $\theta_i$ defines an orientation.
   For all considered models, the orientation $\theta_i$ gives the direction of self-propulsion of the particles.
   The program covers two classes of models.
   In the first class of Vicsek type, the microscopic dynamics is given by certain time discrete rules.
   In the second class, the dynamics is described by overdamped Langevin equations.
   In the following we explain the two classes in detail.
   \subsection{Rule based Vicsek type dynamics}
   In this subsection we consider a discrete time step that consists of two parts: \textit{collision} and \textit{streaming}.
   In the \textit{collision} part all orientations $\theta_i$ are updated.
   In the \textit{streaming} part all positions $(x_i, y_i)$ get updated.
   Thus one time step always consists of first \textit{collision} and second \textit{streaming}.
   The collision update follows the rule:
   \begin{align}
	   \theta_i(t+1)=f(\theta_i(t), \{\theta_j(t)\}_{j\in \Omega_i}) + \eta \cdot \xi_i(t),
	   \label{eq:collision}
   \end{align}
   where the function $f$ specifies the collision rule. 
   It depends on the orientation of particle $i$ and of all of its neighbors before the collision.
   By $\Omega_i$ we denote the set of indexes of neighbors of particle $i$.
   In all considered models, the definition of the neighborhoods only depends on the positions of the particles.
   The considered neighborhood definitions are discussed in subsection \ref{sec:neighborhoods}.
   The last term in \eqref{eq:collision} describes a noise term, where $\xi_i(t)$ are independent random variables that are distributed uniformly on $[-\pi, \pi]$ and $\eta\in[0, 1]$ describes the noise strength. 

   Regarding the collision rule, we consider two models.
   The first is the famous Vicsek Model (VM) \cite{VCBCS95}, where
   \begin{align}
	   f(\theta_i, \{\theta_j\}_{j\in \Omega_i}):= arg\bigg[ exp(i\theta_i) + \sum_{j\in \Omega_i}\exp(i\theta_j) \bigg] + \omega,
	   \label{eq:collision_VM}
   \end{align}
   where the prefactor $i$ in the exponents is the imaginary unit not to be confused with the index $i$. 
   The parameter $\omega$ introduces some chirality on the orientation dynamics.
   Roughly speaking, in the VM dynamics, each particle picks up some kind of an average direction of its neighbors and itself.
   It is then rotated by $\omega$ and disturbed by noise. 
   In \cite{VCBCS95} the model was considered with $\omega=0$.

   We refer to the second considered collision rule as nematic Vicsek model (NVM).
   It was introduced in \cite{CGGP08, GPBC10} with $\omega=0$.
   It is given by
   \begin{align}
	   &f(\theta_i, \{\theta_j\}_{j\in \Omega_i}):= 
	   \notag
	   \\
	   &arg\bigg[ exp(i\theta_i) + \sum_{j\in \Omega_i}\exp[i(\theta_j+\pi/2(1-sign(cos(\theta_i-\theta_j))))] \bigg] + \omega,
	   \label{eq:collision_NVM}
   \end{align}
   where $sign(x)=1$ for $x\ge0$ and $sign(x)=-1$ for $x<0$.
   In this dynamics a particle tries to align with its neighbors that have an orientation that differs by no more than $\pi/2$ and it tries to anti-align with neighbors that have orientations that differ by more than $\pi/2$.

   The streaming part of the update is given by
   \begin{align}
	   x_i(t+1)&=x_i(t)+v \cos(\theta_i(t+1)),
	   \notag
	   \\
	   y_i(t+1)&=y_i(t)+v \sin(\theta_i(t+1)),
	   \label{eq:streaming}
   \end{align}
   where $v$ describes the particle speed.

   \subsection{Overdamped Langevin dynamics}

   For this class of models the dynamics is continuous in time.
   It is given by the following set of differential equations
   \begin{align}
	   &\dot{x}_i=v \cos(\theta_i),
	   \notag
	   \\
	   &\dot{y}_i=v \sin(\theta_i),
	   \notag
	   \\
	   &\dot{\theta}_i= \Gamma h(|\Omega_i|)\sum_{j\in \Omega_i}\sin(o\cdot(\theta_j-\theta_i)  ) + \omega + \eta\cdot \xi_i,
	   \label{eq:Langevin_dynamics}
   \end{align}
   where $v$ denotes the particles speed and $\Omega_i$ the neighbor set of particle $i$ as before.
   The parameter $\Gamma$ defines a coupling strength.
   The function $h$ gives some weight to the interaction depending on the total number of neighbors.
   Common choices are $h(n)\equiv 1$ and $h(n)\equiv 1/(1+n)$, see e.g. \cite{BGHP20, Chate20} for an overview of considered models of this type.
   The program has implemented those two weight functions, but it also allows to specify another arbitrary weight function.
   We refer to the dynamics with $h\equiv 1$ as \textit{additiveL} and with all other weight functions as \textit{nonadditiveL}.
   The integer parameter $o$ defines the order of the interaction.
   For $o=1$ particles tend to align (polar alignment), for $o=2$ particles tend to either align or anti-align (nematic alignment), etc..
   Here, the noise strength $\eta$ is from $[0, \infty]$ and $\xi_i(t)$ denote independent Gaussian white noise terms.
   The equations for the x- and y-positions are ordinary differential equations and the equation for the orientation dynamics is a stochastic differential equation.
   Because the noise is purely additive Ito- and Stratonovic-interpretation of the stochastic differential equation coincide here.
   In the program they are integrated using the Euler-Maruyama-scheme, see e.g. \cite{KP92}.
   \subsection{Neighborhoods\label{sec:neighborhoods}}
   We consider two possible neighborhood definitions: metric neighborhoods and metric free/topological neighborhoods.
   They both have in common that we do not count a particle as its own neighbor (although the particles own orientation effects the collision rule for Vicsek type interactions).

   For metric interactions, for a given particle $i$, all other particles that are closer to particle $i$ than a distance $R$ are considered to be neighbors of particle $i$.
   That means
   \begin{align}
	   \Omega_i:=\{j\neq i:\sqrt{(x_i-x_j)^2+(y_i-y_j)^2}\le R\}.
	   \label{eq:metric_neighborhood}
   \end{align}

   For metric free interaction neighborhoods, that are also called topological interaction neighborhoods, each particle has a fixed number of (closest) neighbors that is denoted by $k$.
   In that case the set $\Omega_i$ consists of the indexes of the $k$ particles, different from particle $i$, that are closest to particle $i$.

   We refer to the metric free neighborhood by the abbreviation 'mf'.
   Whenever we do not specify anything else, we refer to the metric neighborhood definition.
   Thus we might refer to the metric free Vicsek model as 'mfVM' and to the standard Vicsek model as 'VM', etc..
   The weight function introduced in the overdamped Langevin dynamics makes no sense in the metric free case because the number of neighbors is the same for all particles and the corresponding weight can be absorbed in the interaction strength $\Gamma$.
   Thus, we always use $h\equiv 1$ in the metric free case and refer to the corresponding overdamped Langevin dynamics as 'mfL'.
   \subsection{Boundary Conditions}
   The program has implemented periodic and reflecting boundary conditions.
   They can be specified for x- and y-direction separately.
   Here we will explain them only for the x-direction.

   For periodic boundary conditions we introduce two virtual image particles for each real particle by shifting the x-coordinate by $\pm L_x$.
   We only calculate the time evolution of the real particles, however, they interact also with the virtual image particles.
   When a particle leaves the simulation box $[0, L_x]$ during the dynamics, it is set back into the box by either adding or subtracting $L_x$.

   For reflecting boundary conditions, for each real particle, there are introduced two virtual image particles as well.
   However, in this case, they are created by mirroring position as well as orientation at the lines $x=0$ and $x=L_x$.
   For the orientation that means that $\theta \rightarrow \pi - \theta$.
   Particles take into account virtual image particles in their interactions in the same way as for periodic boundary conditions.
   When a particle leaves the simulation box $[0, L_x]$ during the dynamics it is set back into it by applying the mirroring of position and orientation at either the line $x=0$ or $x=L_x$. That means $\theta \rightarrow \pi - \theta$ and either $x\rightarrow -x$ or $x\rightarrow -x + 2L_x$.

   Note that the virtual image particles have bean just introduced for illustration. 
   It is not necessary to produce such image particles in the implementation in order to calculate the dynamics of the real particles.

   \subsection{Distinguished Particle Types/Species}

   The dynamics described above depends on a number of parameters.
   The program allows to use different values for speed $v$, noise strength $\eta$, chirality $\omega$ and coupling strength $\Gamma$ for different particles.
   More precisely, it allows to consider e.g. two different species of particles, say A- and B-particles, such that the dynamics of A-particles uses $v_A, \eta_A, \omega_A, \Gamma_{AA}$ and $\Gamma_{AB}$ and the dynamics of B-particles uses the parameters $v_B, \eta_B, \omega_B, \Gamma_{BB}$ and $\Gamma_{BA}$.
   Two-species models that are implemented in the program have been studied e.g. in \cite{Menzel12} (varying coupling) or \cite{VCMS21} (varying chirality).
   The number of A-particles $N_A$ can be different from the number of B-particles $N_B$.
   There can be an arbitrary number of particle species.
   However, if the number of species is very large (in the extreme case =$N$ such that each particle has a different parameter) the program is extremely inefficient.
   The current implementation is designed for a small number of species only.

   \section{Implementation\label{sec:implementation}}

   \subsection{Time Evolution}
   The major part in the implementation is the determination of the neighborhoods.
   The collision and streaming rules as well as the Euler-Maruyama discretization of the stochastic differential equation are straight forward.
   It should be mentioned that the glib-implementation of the Mersenne-Twister-algorithm is used to generate uniform pseudo random numbers.
   For the Euler-Maruyama scheme, Gaussian random numbers are produced by the Box-Muller-algorithm.
   In the following we describe the determination of metric and metric free neighborhoods.

   \subsection{Metric Neighborhoods}

   Cell lists are used to determine the neighborhoods.
   The cell size is close to the interaction radius $R$ in both dimensions.
   It is a little larger than $R$ in order to guarantee a perfect tiling of the simulation box.
   In each step, for each cell, a list is produced that contains all particles that lie inside the cell.
   To find the neighbors of particle $i$, one needs to check the distance to all particles that are in the same cell as particle $i$ or in the eight surrounding cells.
   So, one needs to check all particles from nine boxes in total.
   Thus, for homogeneous systems with density $\rho_0=N/L_x/L_y$ the complexity of the algorithm is $N \rho_0 9 R^2$.
   Homogeneous states are somehow the best possible scenario.
   In many cases, e.g. for high density Vicsek bands, particles accumulate locally.
   If this happens, the simulations usually slow down a bit.
   However, the complexity remains proportional to the particle number $N$ as long as the system exhibits a well-defined thermodynamic limit.

   \subsection{Metric free Neighborhoods}

   For the metric free neighborhood, an effective radius is defined as
   \begin{align}
	   R_{eff}=\sqrt{L_x L_y k /N/\pi},
	   \label{eq:effective_radius}
   \end{align}
   where $k$ is the number of neighbors.
   The effective radius is then used to define cells as in the metric case and produce cell lists.
   For each particle, a neighbor list with the next $k$ neighbors is produced by first checking all particles from the box of particle $i$.
   If the closest point from the remaining boxes is closer than the $k$th neighbor also the surrounding eight boxes will be checked.
   In the next step the surrounding 16 boxes will be checked and so on.
   The exact complexity of the algorithm depends on the fluctuations of the local density, however, also in this case it is proportional to the total particle number $N$.

   \subsection{Measurements\label{subsec:measurements}}
   The dynamics of the system can be iterated with or without performing any measurements.
   If measurements are done on the fly the following quantities are measured
   \begin{itemize}
	   \item Histogram of the orientations $\theta$.
	   \item Histogram of the number of neighbors $|\Omega_j|$.
	   \item Polar order parameter $p:=|\frac{1}{N}\sum_{j=1}^N\exp(i\theta_j)|$ and its first four moments.
	   \item Nematic order parameter $q:=|\frac{1}{N}\sum_{j=1}^N\exp(2i\theta_j)|$ and its first four moments.
	   \item The first four moments of the ensemble averaged number of neighbors $\bar{n}=\frac{1}{N}\sum_{j=1}^N|\Omega_j|$. 
   \end{itemize}
   All histograms and moments are stored internally unnormalized, however, when they are accessed they are returned normalized.
   If a measurement is performed for $n$ steps, the program performs $n$ updates, such that there are in total $n+1$ states from $t=0$ to $t=n$.
   The measurements and all the statistical analysis behind only uses $n$ of those $n+1$ steps, namely from $t=0$ to $t=n-1$ (the last step is not used for the statistics).

   \subsection{Memory Usage}
   On usual machines where a C double takes $8$ Bytes a simulation object takes $40\times N$ Bytes $+N\times$the size of a pointer (8Bytes on 64bit system, 4Bytes on 32bit system) $+$the number of cells$\times$ the size of a pointer $+$some small overhead for storing parameters and measurement results.
   Note that currently memory for both cell list: for metric and metric free neighborhoods is allocated.
   If memory is an issue, for metric neighborhood simulation one can set the parameter $k$ (which is not used) to a very large value leading to a large $R_{eff}$.
   For metric free simulations one can set $R$ (which is not used) to a very large value.
   In that way one can minimize the memory usage of the unused cell list.

   \subsection{Disk Space Usage}
   It is possible to save the full state of the simulation to disk in binary format and read it in later for analysis or in order to continue the simulation.
   The system state including the particle configurations, all measurement results and the state of the pseudo random number generator is fully saved.
   Hence an intermediate save/load operation leads to identical results as a simulation without interim saving/loading.
   Saving is, however, not strictly type save as in principle the size of C double depends on the system.
   Assuming usual machines that use 8 Bytes for a C double, saving a simulation object takes $24\times N$Bytes $+$some small overhead to save a simulation object.

   \section{Usage\label{sec:usage}}
   Details on the usage of the package can be found in the documentation (either in the README or by calling help(aappp)).
   Here, we only give some general remarks.
   Independent on the model that will be simulated, a simulation object is always initialized by the function \textit{aappp\_init}.
   Not all parameters have to be given.
   For parameters that are not given, default values are used.
   They can be found in the documentation.
   Possible parameters are: velocity $v$ as \textit{v=\ldots}, interaction radius $R$ as \textit{R=\ldots}, noise strength $\eta$ as \textit{eta=\ldots}, chirality $\omega$ as \textit{omega=\ldots}, coupling $\Gamma$ as \textit{gamma=\ldots}, simulation box size in x-direction $L_x$ as \textit{Lx=\ldots}, simulation box size in y-direction $L_y$ as \textit{Ly=\ldots}, step size of Euler-Maruyama-scheme $\Delta t$ as \textit{dt=\ldots}, neighbor number for metric free models $k$ as \textit{kn=\ldots}, interaction order for overdamped Langevin dynamics $o$ as \textit{order=\ldots}, boundary condition in x-direction (periodic if=0, reflecting if=1) bx as \textit{bx=\ldots}, boundary condition in y-direction by as \textit{by=\ldots}, particle number $N$ as \textit{N=\ldots}, seed for pseudo random number generator as \textit{seed=\ldots}, weight function $f$ for nonadditive overdamped Langevin dynamics as \textit{weight\_function=\ldots} (if no weight function is specified $f(n)=1/(1+n)$ is used for all $n$), weight vector length as \textit{weight\_vector\_length=\ldots} (the weight function is called only once for $n=0, \cdots, weight\_vector\_length-1$ and the results are saved in memory, if the number of neighbors is larger than weight\_vector\_length-1 in some case, then the weight $f(weight\_vector\_length-1)$ is used), the number of bins used for the orientation histogram as \textit{binnum\_theta=\ldots}, the number of bins used for the number of neighbor histogram as \textit{binnum\_neighbors=\ldots}.
   In the current implementation, reflecting boundary conditions are used whenever $bx/by\neq 0$, however, $bx/by=1$ should be used as other values might be implemented later.
   When different species of particles are simulated, the following parameters can be lists: v, eta, omega, N and gamma can be a list of lists (e.g. a coupling matrix). 

   Time evolution is done by the functions\\ \textit{*model*\_update\_timesteps(simulation, timesteps)}, where \textit{*model*} should be replaced by the abbreviation for the model that should be iterated as e.g. \textit{mfVM}, simulation is the simulation object initialized before and timesteps is the number of time steps that will be iterated.

   Similarly, time evolution with measurements is done by\\ \textit{*model*\_measurement\_timesteps(simulation, measurement)}.
   The results of measurements can be obtained by \textit{aappp\_get\_results(simulation)}.

   A simulation object can be saved to or loaded from a file on disc via \textit{aappp\_save(simulation, 'filename')}, \textit{simulation=aappp\_load('filename')}.

   \section{Numerical Results\label{sec:examples}}
   \subsection{Comparison to Literature}
\begin{figure}[h]
	   \begin{center}
	   \includegraphics[width=0.32\textwidth]{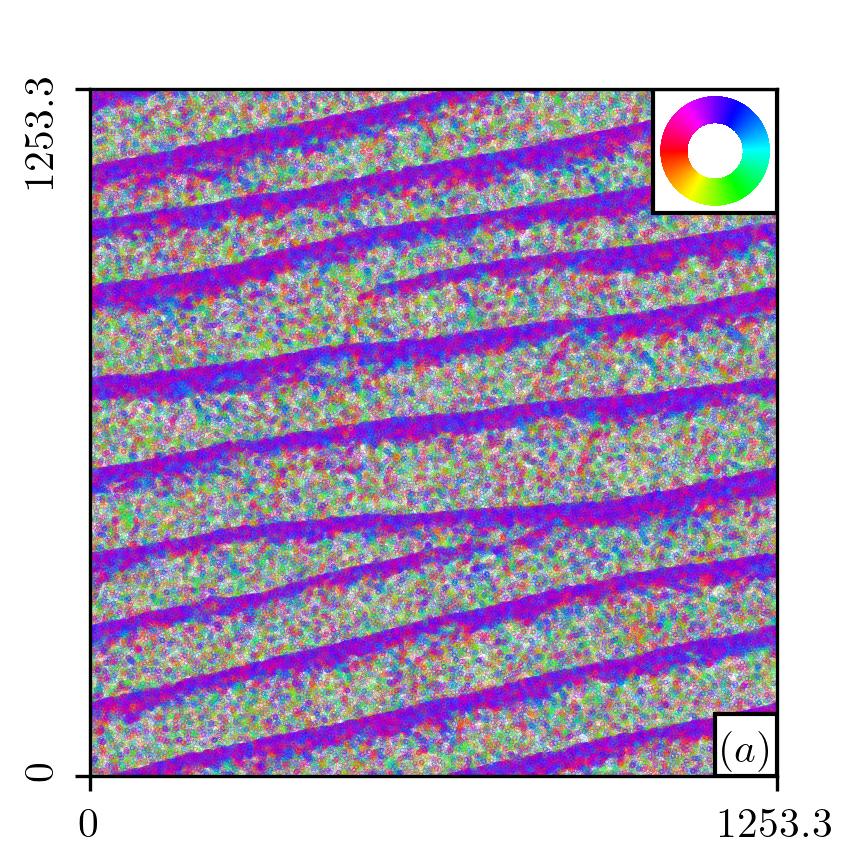}
	   \includegraphics[width=0.32\textwidth]{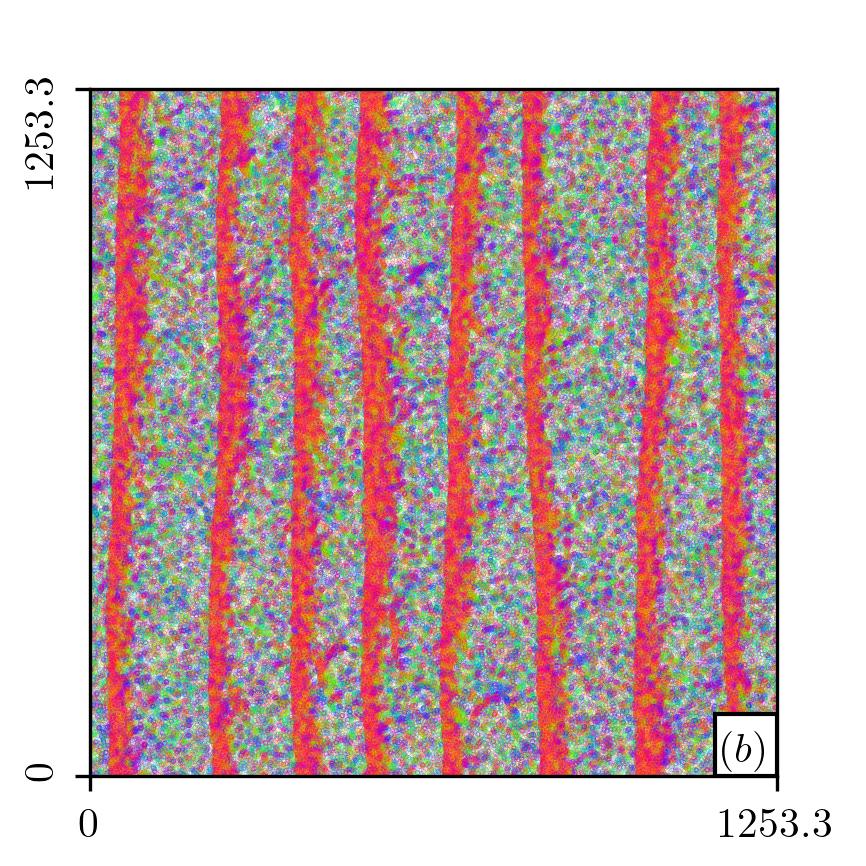}
	   \includegraphics[width=0.32\textwidth]{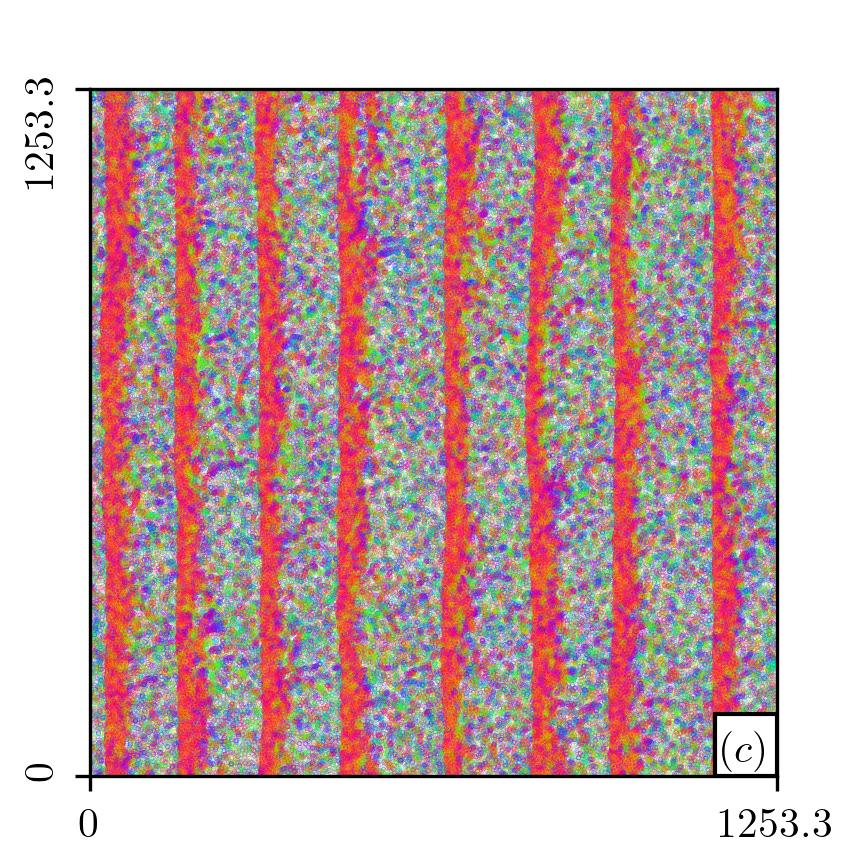}\\
	   \includegraphics[width=0.32\textwidth]{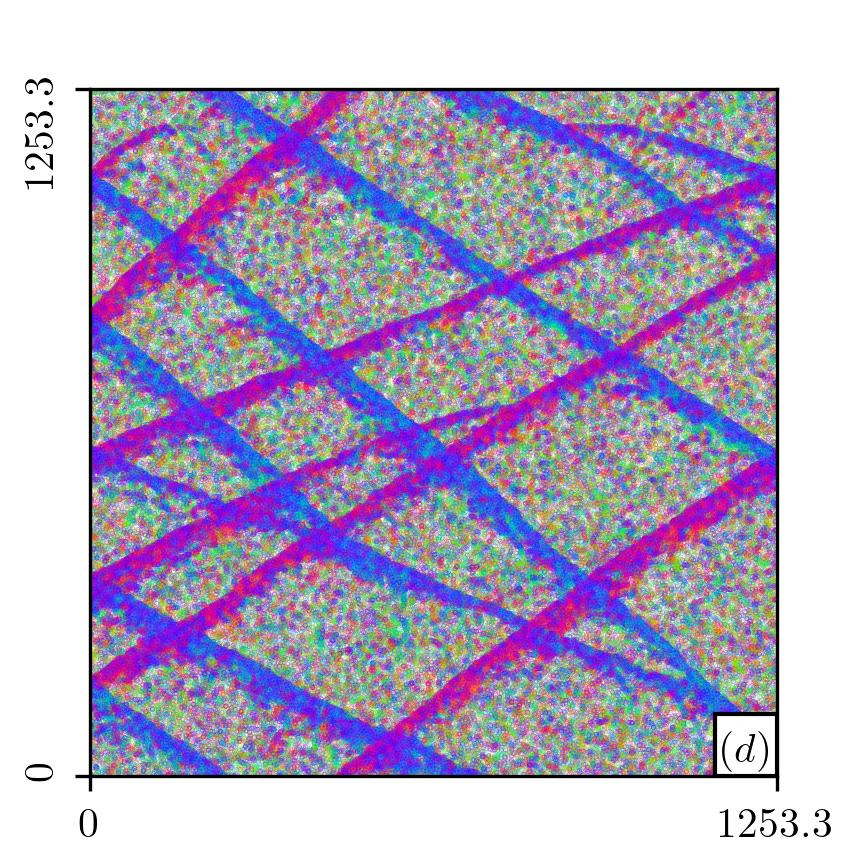}
	   \includegraphics[width=0.32\textwidth]{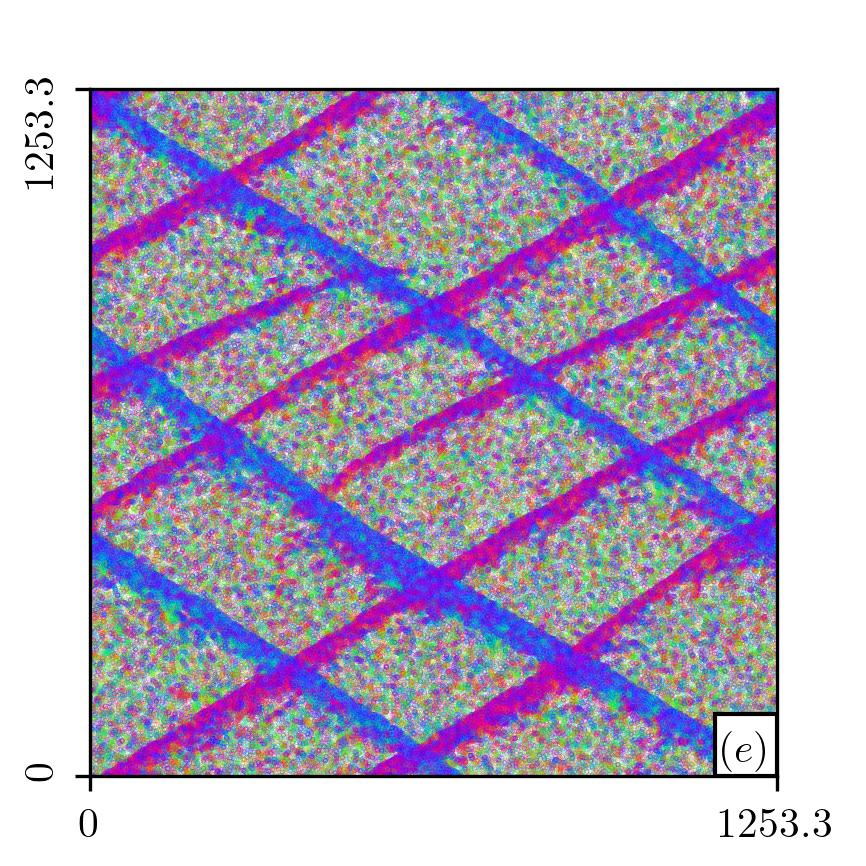}
	   \end{center}
	   \caption{Snapshots of the standard Vicsek model with $N=10^6$ particles.
	   Physical parameters are $R=1$, $v=1$, $\eta=0.37$ and $\omega=0$.
	   For $(a), (b)$ and $(d)$ uniform, isotropic random initial conditions have been used.
	   For $(c)$ and $(e)$ the final states of $(b)$ and $(d)$ have been used as initial conditions, respectively.
	   Simulations $(a), (c)$ and $(e)$ use periodic boundary conditions in both directions.
	   Simulations $(b)$ and $(d)$ use reflecting boundary conditions in $x$- and $y$-direction, respectively.
	   Snapshot $(a)$ was taken after $10^5$ time steps, snapshots $(b)-(e)$ after $5\times 10^4$ time steps.
	   For each particle a point was drawn at its position. The color encodes the orientation of the particle according to the color wheel shown in $(a)$}
	   \label{fig:VM}
   \end{figure}

   We perform a few test simulations and compare them to published results of some of the models that can be handled by the program.
   Simulation data discussed here are available at \cite{data}.
   First, we look at simulations of the famous standard Vicsek model (\textit{VM}).
   In Fig.~\ref{fig:VM} we display simulation results of a parameter set that was used in Fig.~1 $(iii)$ of Ref. \cite{KI20}.
   In Fig.~\ref{fig:VM} $(a)$ we see a snapshot after $10^5$ time steps.
   We observe a nice band pattern, very similar to Fig.~1 $(iii)$ of Ref. \cite{KI20}.
   There are still two defects in the pattern that are most likely vanishing in longer simulations.

   We remark that in the experience of the author, in a typical situation, the bands that arise from random initial conditions move in an arbitrary direction (that is in most cases not equal to the $x$- or the $y-$axis direction, see also supplemental material of \cite{KI20}).
   However, in many cases, bands that are moving nicely in the direction of the coordinate axis are presented without commenting on it.
   In the authors experience it is not so easy to create such directed band patterns.
   One might try to initialize all particles moving e.g. in the $x$-direction.
   However, then usually initial polar order gets lost in the beginning of the simulation and later, polar order builds up again spontaneously in (most likely) some other direction.

   Here, we want to present a simple recipe to (artificially) create band patterns that are aligned with one of the coordinate axis whenever this is desired.
   This works by applying reflecting boundary conditions in one of the directions and thus stopping a directed motion into this direction.
   Reflecting boundary conditions in $y-$direction have been used in Fig.~\ref{fig:VM} $(b)$ leading to bands that are nicely moving into $x$-direction.
   Once such a pattern is created one can switch boundary conditions back to periodic and the pattern survives, see Fig.~\ref{fig:VM} $(c)$.

   Apparently this method does not alway succeed.
   In Fig.~\ref{fig:VM} $(d)$ we apply periodic boundary conditions in $x$-direction and we observe that the bands get really reflected at the boundaries given by $x=0$ and $x=L_x$.
   After switching back to periodic boundary conditions we observe a cross sea pattern, see Fig.~\ref{fig:VM} $(e)$ and cf. \cite{KI20}.
   For the given parameter set this phase is most likely not stable and collapsing after some very long time.
   In order to enforce the direction of the bands to be aligned with one of the coordinate axis one might switch between periodic and reflecting boundary conditions with a higher frequency, however, in order to keep the discussion short we do not show results of such a simulation protocol here.

\begin{figure}[h]
	   \begin{center}
	   \includegraphics[width=0.32\textwidth]{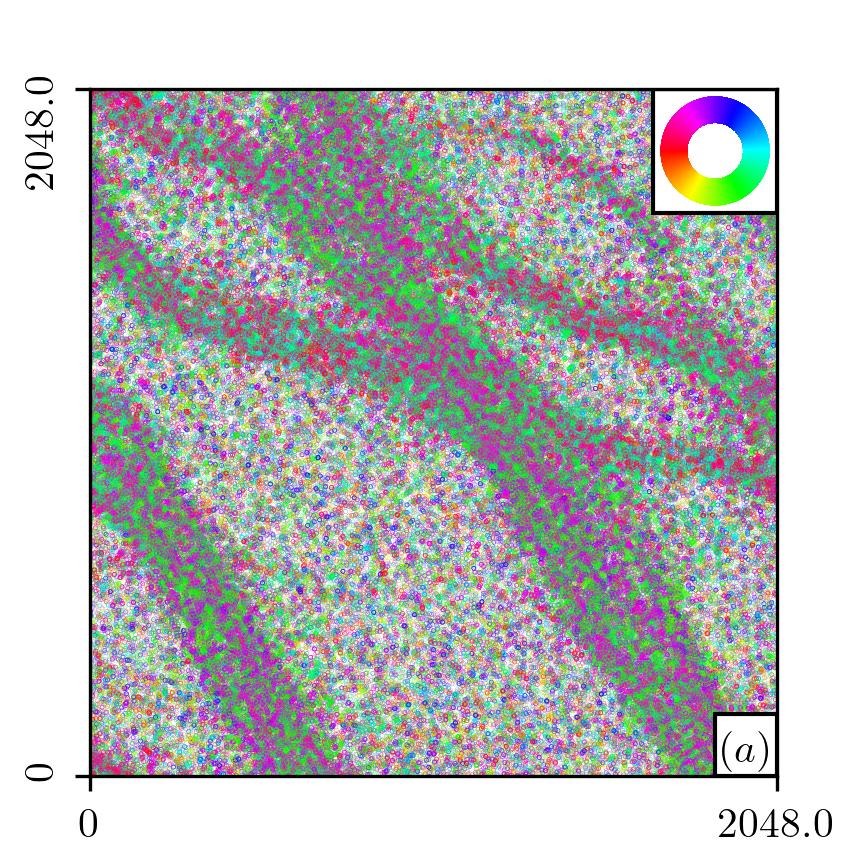}
	   \includegraphics[width=0.32\textwidth]{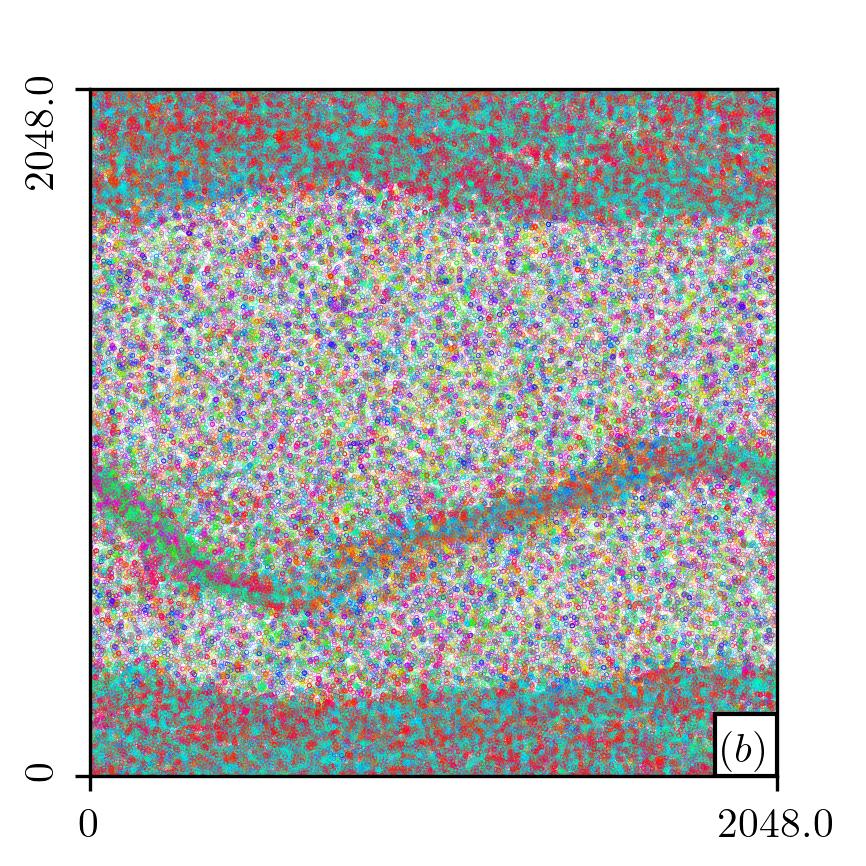}
	   \includegraphics[width=0.32\textwidth]{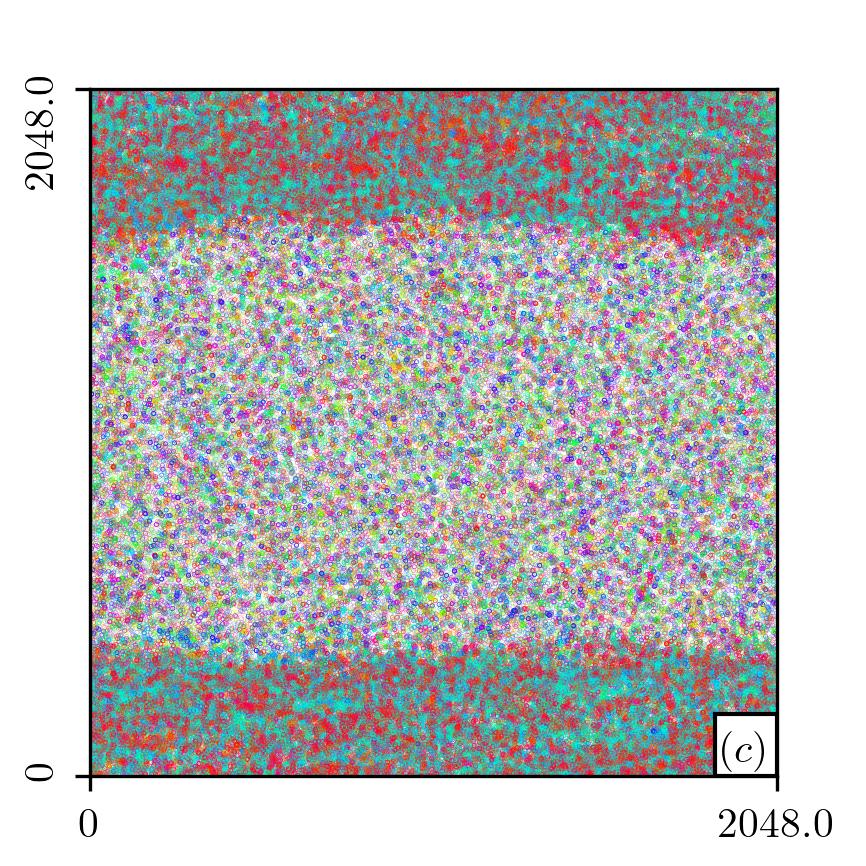}
	   \end{center}
	   \caption{Snapshots of the nematic Vicsek model (NVM) with $N=524288$ particles. Physical parameters are $R=1$, $v=0.5$, $\eta=0.065$ and $\omega=0$.
	   Uniform and isotropic random initial conditions have been used for $(a)$ and $(b)$.
	   For $(c)$, the final state of $(b)$ was used as initial condition.
	   For $(a)$ and $(c)$ periodic boundary conditions have been used in both directions.
	   For $(b)$ reflecting boundary conditions have been used in $y$-direction.
	   All snapshots have been taken after $10^5$ time steps.
	   Plotting was done as in Fig.~\ref{fig:VM}}
	   \label{fig:NVM}
   \end{figure}

   Next, we perform simulations of the nematic Vicsek model (\textit{NVM}) using a parameter set from Fig.~2 $(c)$ of \cite{GPBC10}.
   From uniform, isotropic random initial conditions we obtain after $10^5$ steps a not yet perfectly ordered, nematic band structure that is directed somehow diagonal, see Fig.~\ref{fig:NVM} $(a)$.

   In order to speed things up and obtain a nice band oriented in $x$-direction we employ the same method as for the Vicsek model.
   We first use reflecting boundary conditions in $y$-direction for a few time steps, see Fig.~\ref{fig:NVM} $(b)$ and switch to periodic boundary conditions later.
   Eventually, we arrive at a state that looks like Fig.~2 $(c)$ of Ref. \cite{GPBC10}, see Fig.~\ref{fig:NVM} $(c)$.

\begin{figure}[h]
	   \begin{center}
	   \includegraphics[width=0.48\textwidth]{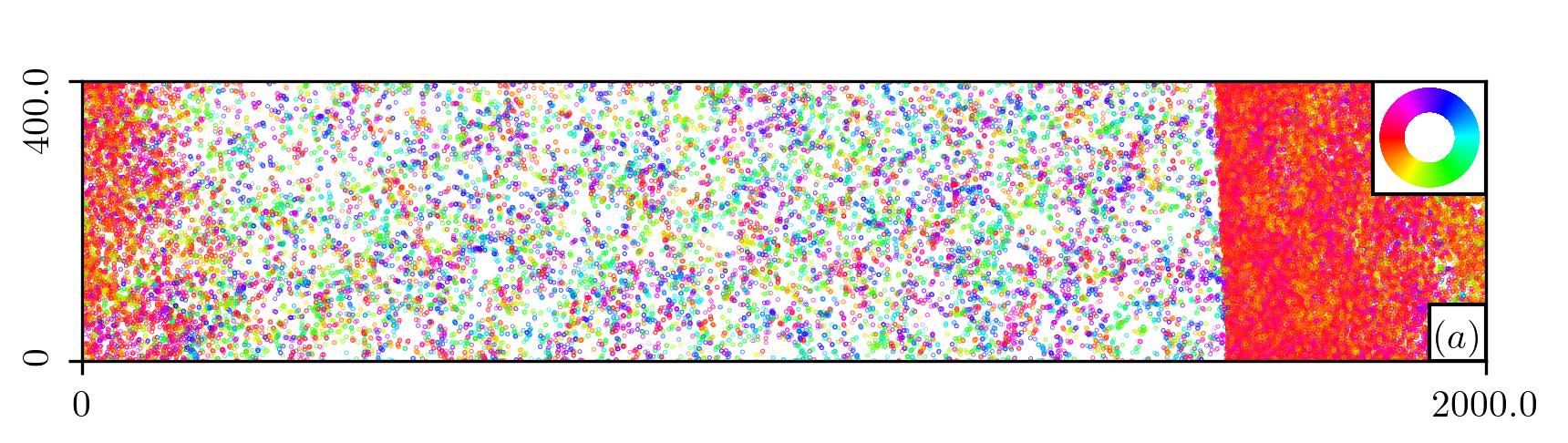}
	   \includegraphics[width=0.48\textwidth]{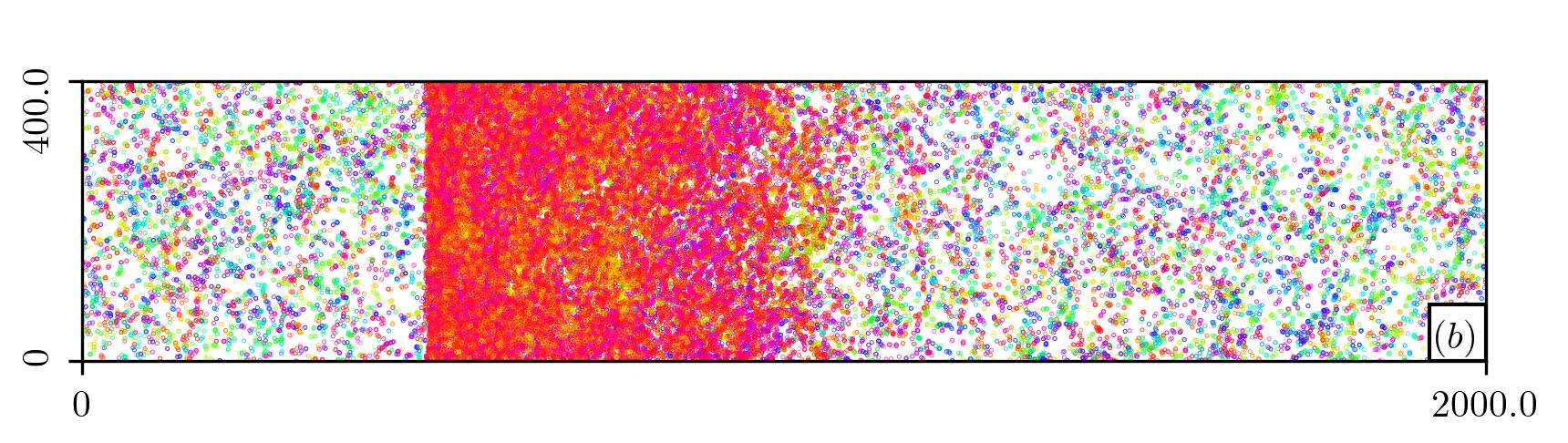}\\
	   \includegraphics[width=0.48\textwidth]{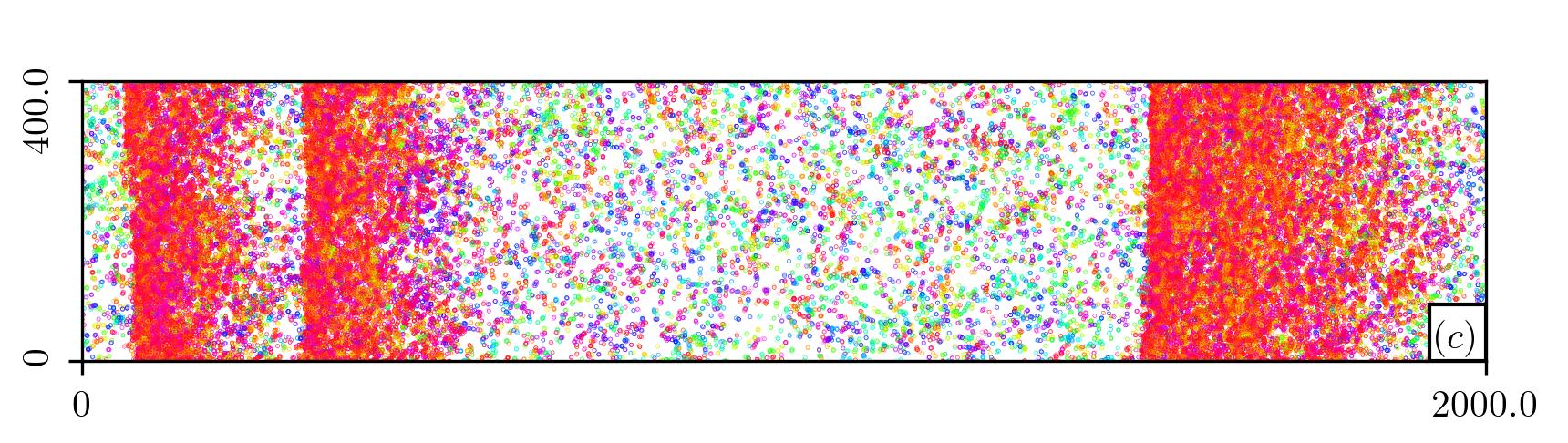}
	   \includegraphics[width=0.48\textwidth]{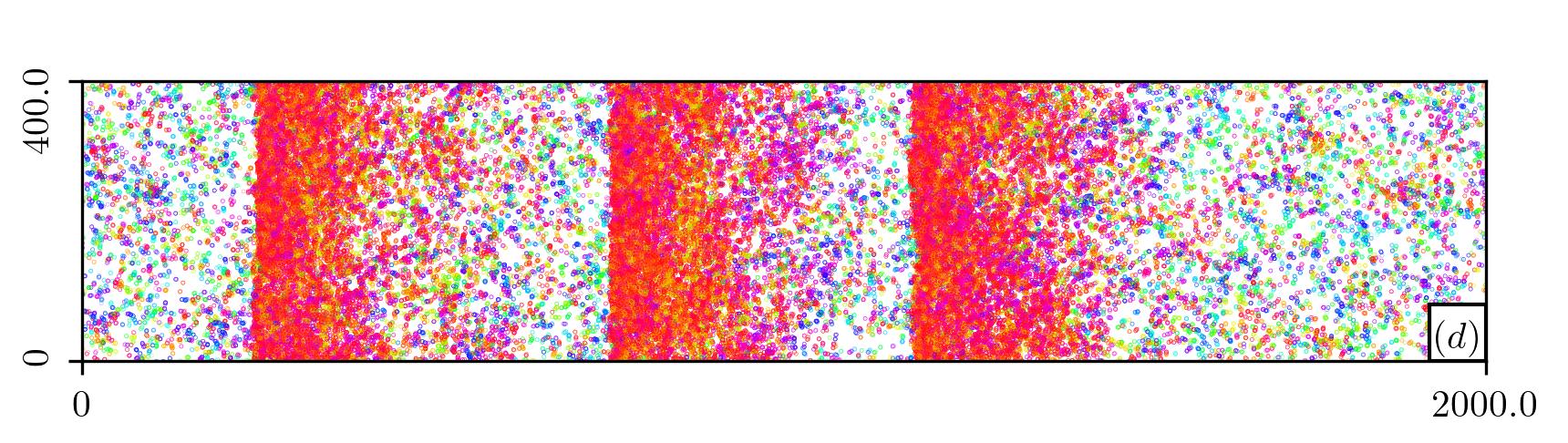}
	   \end{center}
	   \caption{Snapshots of the metric free Vicsek model (mfVM) for $N=2\times 10^5$ ($(a)$ and $(b)$), and $N=3.2\times 10^5$ ($c$ and $(d)$) particles.
	   Physical parameters are $k=2$, $v=0.2$, $\eta=0.08$ and $\omega=0$.
	   Thus, each particle interacts with itself and the next two neighbors.
	   Uniform and isotropic random initial conditions have been used in $(a)$ and $(c)$.
	   The final states of $(a)$ and $(c)$ have been used as initial conditions for $(b)$ and $(d)$, respectively.
	   For $(a)$ and $(c)$ periodic boundary conditions in $x$-direction and reflecting boundary conditions in $y$-direction have been used.
	   For $(b)$ and $(d)$ periodic boundary conditions have been used in both directions.
	   Snapshots $(a)$ and $b$ have been taken after $5\times 10^4$ time steps.
	   Snapshot $(c)$ was taken after $2.5\times 10^5$ time steps and snapshot $(d)$ was taken after $8\times 10^5$ time steps.
	   Plotting was don as in Fig.~\ref{fig:VM}.}
	   \label{fig:mfVM}
   \end{figure}

   Until recently, it was believed that the flocking state for metric free models is spatially homogeneous, cf. e.g. \cite{Chate20}.
   However, it was found that polar ordered bands exist also there, see \cite{MCNSTW21}.
   We performed simulations of the metric free Vicsek model (\textit{mfVM}) with parameters as in Fig. 3 of \cite{MCNSTW21}.
   We used again the same method as before, switching between reflecting and periodic boundary conditions in order to align the resulting pattern with the $x$-axis.
   The final states of the simulations seem to be very close to the ones presented in Ref. \cite{MCNSTW21}, cf. Fig.~\ref{fig:mfVM} $(b)$ and $(d)$.

\begin{figure}[h]
	   \begin{center}
	   \includegraphics[width=0.40\textwidth]{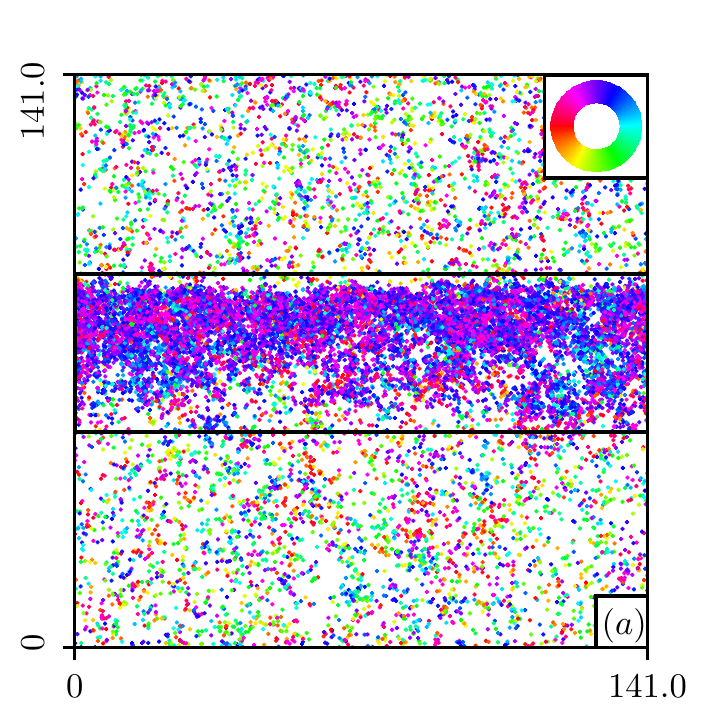}
	   \includegraphics[width=0.40\textwidth]{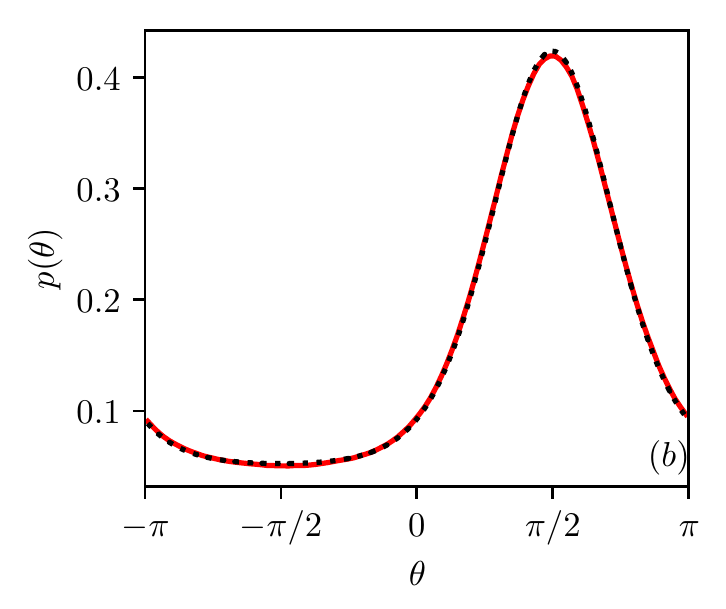}
	   \end{center}
	   \caption{Simulation results of nonadditive overdamped Langevin dynamics (nonadditiveL) with weight function $h(n)=1/(n+1)$ for $N=19881$ particles. 
	   Physical parameters are $R=1$, $v=1$, $\eta=1/\sqrt{2}$, $\Gamma=1$ and $\omega=0$.
	   The integration step size is $\Delta t = 0.003$.
	   The simulation was started from uniform and isotropic random initial conditions.
	   Periodic boundary conditions have been used.
	   The snapshot $(a)$ was taken after $101\times10^4$ time steps.
	   The orientation distribution $(b)$ was measured for $10^4$ time steps after thermalizing for $10^6$ time steps.
	   The red line shows the normalized histogram from the simulation.
	   The dotted black line shows a superposition of a von Mises distribution with weight $\alpha$ and a uniform distribution with weight $1-\alpha$.
	   The weight $\alpha$ was determined as the fraction of particles that are between the two horizontal black lines in $(a)$, $\alpha \approx 0.703$.
	   Those lines have been set by eye in order to mark the band of polarly ordered particles.}
	   \label{fig:nonadditiveL}
   \end{figure}

   For an overdamped Langevin model with weight function $h(n)=1/(n+1)$ (\textit{nonadditiveL}) we perform a simulation for a parameter set as in Fig. 1 $(a)$ of Ref. \cite{CSP21}, middle snapshot.
   We observe a very similar snapshot, see Fig.~\ref{fig:nonadditiveL} $(a)$.
   Here, we also display the measured orientation distribution, see Fig.~\ref{fig:nonadditiveL} $(b)$, because there is some analytical theory to compare with.
   It is known that the steady state orientational distribution of the considered model is a von Mises distribution within homogeneous mean field theory \cite{CSP21, KI21}.
   That means the orientation distribution is of the type $p(\theta)=\exp[K\cos(\theta-\theta_0)]/Z$, where $K$ and $\theta_0$ are distribution parameters and $Z$ is a normalization factor.

   It is well known and we can easily observe from Fig.~\ref{fig:nonadditiveL} $(a)$ that the density is not homogeneous.
   Instead there is micro phase separation.
   Within a high density band the particles are polarly ordered while the particles in the surrounding low density region are disordered, cf. also \cite{SCT15}.
   For that reason we have identified the high density band by eye in Fig.~\ref{fig:nonadditiveL} $(a)$ as the region between the two horizontal black lines.
   We count the number of particles in that region and denote its fraction from all particles by $\alpha\approx 0.703$.

   We use the von Mises ansatz only for the fraction $\alpha$ of particles in the band and assume that the fraction $1-\alpha$ of particles outside the band is isotropic oriented.
   Eventually we obtain the von Mises parameters $K$ and $\theta_0$ from the expectation values $\langle \cos(\theta) \rangle$, $\langle \sin(\theta) \rangle$, which we calculate from the measured distribution $p(\theta)$.
   From the combined von Mises/disordered ansatz we obtain the orientation distribution displayed by the dotted black line in Fig.~\ref{fig:nonadditiveL} which agrees very well with the measured distribution. 

   In a recent study \cite{ZHR22}, it was reported that the metric free overdamped Langevin model exhibits not only one but two phase transitions.
   For small coupling there is disorder as expected.
   For larger coupling there is a transition towards polar order.
   Interestingly, for even larger coupling, there is another transition towards disorder \cite{ZHR22}.
   Disorder at large coupling appears because particles arrange into locally high ordered clusters that only rarely interact with, and thus do not align with other clusters.
   The interaction term used in Ref. \cite{ZHR22} is linear and discontinuous and thus not exactly the same as implemented here (which is harmonic and continuous).
   However, we expect to observe the same qualitative behavior.

\begin{figure}[h]
	   \begin{center}
	   \includegraphics[width=0.40\textwidth]{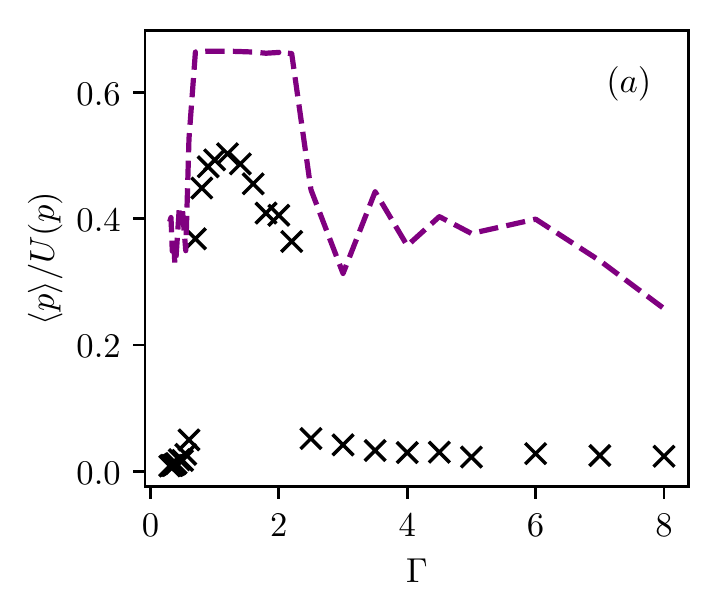}
	   \includegraphics[width=0.40\textwidth]{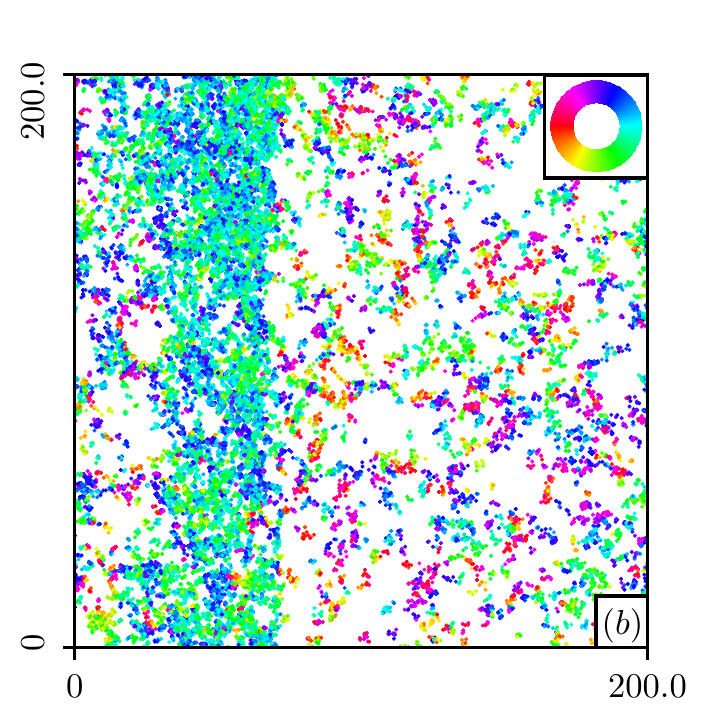}
	   \end{center}
	   \caption{Simulation results of metric free overdamped Langevin model (mfL) for $N=4\times 10^4$ particles.
	   Physical parameters are $k=3$, $v=1.5$, $\omega=0$ and $\Gamma \in \{0.3,\allowbreak 0.32,\allowbreak 0.34,\allowbreak 0.36, 0.38, 0.4, 0.45, 0.5, 0.55, 0.6, 0.7, 0.8, 0.9, 1.0, 1.2, 1.4, 1.6, 1.8, 2.0, 2.2,\allowbreak 2.5,\allowbreak 3,\allowbreak 3.5,\allowbreak 4,\allowbreak 4.5, 5, 6, 7, 8\}$.
	   Thus each particle interacts with its (different from itself) three nearest neighbors.
	   The integration step used is $\Delta t=0.01$.
	   Uniform, isotropic random initial conditions and periodic boundary conditions have been used.
	   The system was thermalized for $10^6$ time steps.
	   Afterwards, moments of the polar order parameter have been sampled for $10^5$ time steps.
	   In $(a)$ we display the first moment of the polar order parameter (black crosses) and the Binder cumulant $U(p)=1-\langle p^4 \rangle/3/\langle p ^2\rangle^2$ (dashed purple line).
	   Only one realization was used for each data point.
	   In $(b)$ we display a snapshot at the end of the measurement from within the ordered phase at $\Gamma=1.6$.}
	   \label{fig:mfL}
   \end{figure}

   As another test of the program we run simulations of the metric free overdamped Langevin model (\textit{mfL}) and measure polar order as a function of coupling.
   Indeed, we find both transitions reported in \cite{ZHR22}.
   In Fig.~\ref{fig:mfL} we display the average polar order parameter and the Binder cumulant of the polar order parameter.
   We have only used one realization for each coupling and simulated only one system size.
   Thus, the measured Binder cumulant fluctuates a bit, but it still shows the characteristic drop from $2/3$ to a smaller value (that should be zero in the thermodynamic limit) indicating two continuous transitions from disorder to order to disorder.

   We also display a snapshot from within the ordered phase (at $\Gamma=1.6$) that shows a band, similar to the metric free Vicsek case, see Fig.~\ref{fig:mfL} $(b)$.
   The presence of such bands was already reported in \cite{ZHR22}.

\begin{figure}[h]
	   \begin{center}
	   \includegraphics[width=0.40\textwidth]{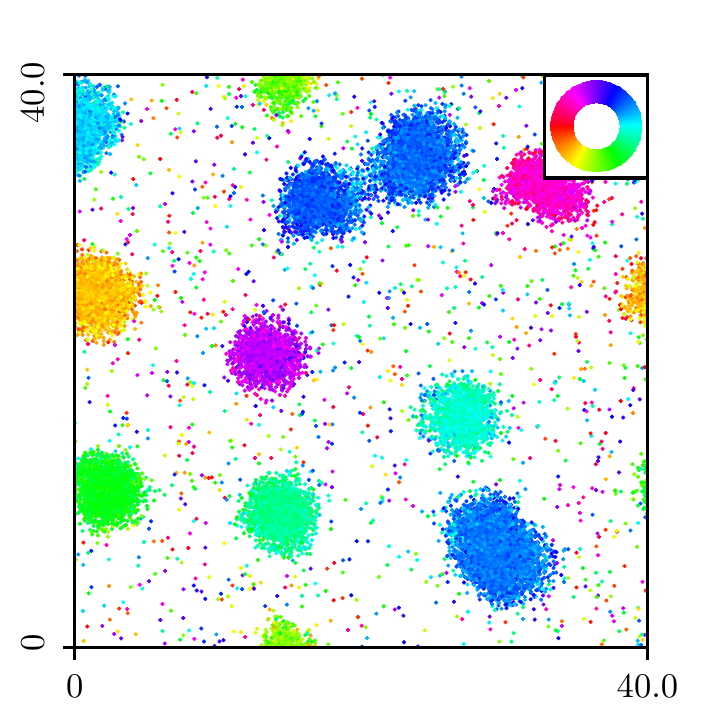}
	   \end{center}
	   \caption{Snapshot of a simulation of additive overdamped Langevin model (additiveL) with chirality for $N=3.2\times 10^4$ particles.
	   Physical parameters are $R=1$, $v=0.5$, $\omega=1.5$, $\eta=1$ and $\Gamma = 0.07$.
	   The integration step used is $\Delta t=0.01$.
	   Uniform, isotropic random initial conditions and periodic boundary conditions have been used.
	   The snapshot was taken after $4.1 \times 10^6$ time steps.
	   }
	   \label{fig:addLchiral}
   \end{figure}

   As the last test we performe a simulation of the additive overdamped Langevin dynamics (\textit{additiveL}) with nonzero chirality. 
   We use the physical parameters of Fig.~2$(d)$ of Ref.~\cite{LL17}.
   A snapshot of the final simulation state is shown in Fig.~\ref{fig:addLchiral}.
   We find the same number and about the same size of polar ordered droplets as in Fig.~2$(d)$ of Ref.~\cite{LL17}.
   \subsection{Performance}
   In this subsection we show the run time of the program for the standard Vicsek model (VM) as a function of particle number $N$ and density $\rho_0$ within the disordered phase.
   All simulations have been performed using a single cpu-core of the \textit{Brain-Cluster} at Universitätsrechenzentrum Greifswald.

\begin{figure}[h]
	   \begin{center}
	   \includegraphics[width=0.45\textwidth]{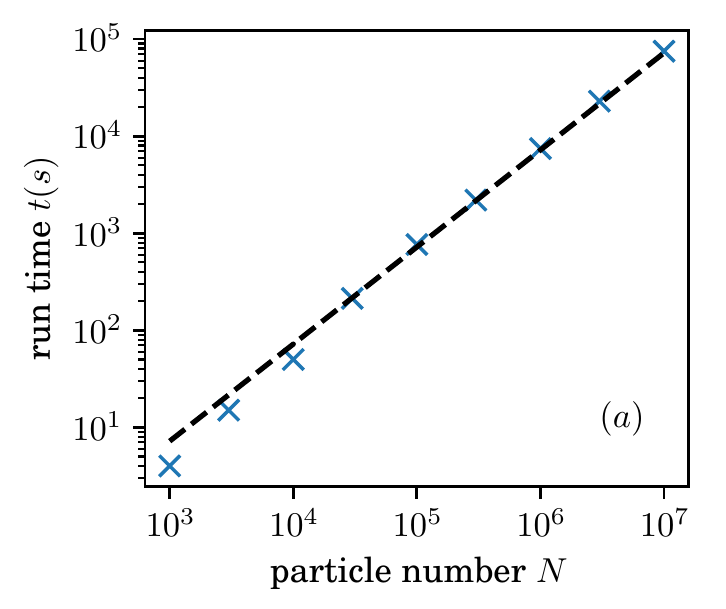}
	   \includegraphics[width=0.45\textwidth]{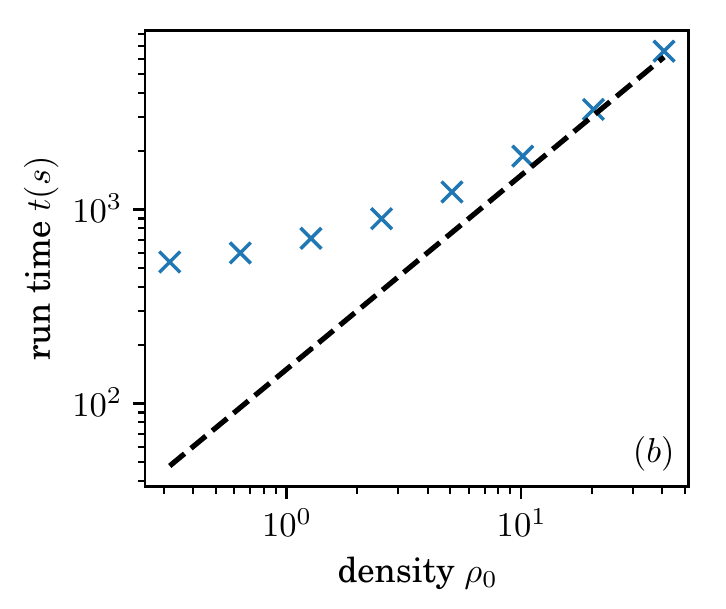}
	   \end{center}
	   \caption{Run time (in $s$) of program initialization with uniform, isotropic random initial conditions $+$ iteration of $10^4$ time steps of the standard Vicsek model (VM) as a function of particle number $N$ $(a)$ and of particle density $\rho_0$ $(b)$. System parameters are $R=1$, $v=1$ and $(a)$: $\eta=0.45$, $\rho_0=0.6366$, $(b)$: $\eta=0.85$, $N=10^5$.
	   The system remains in the disordered state for all simulations.
	   The dashed black line in $(a)$ shows the line $t=0.0072 s \times N$, in $(b)$ it shows the line $t=150s \times \rho_0$, they serve as a guide to the eye.}
	   \label{fig:runtime}
   \end{figure}

   In Fig.~\ref{fig:runtime} $(a)$ we display the run time to perform $10^4$ Vicsek time steps as a function of the particle number at constant global density.
   Parameters have been chosen such that the system is within the disordered phase where the spatial particle density is homogeneous.
   For large particle number, the program run time is proportional to the particle number as argued in Sec.~\ref{sec:implementation}.

   In Fig~\ref{fig:runtime} $(b)$ we display the run time to perform $10^4$ Vicsek time steps as a function of global particle density for constant particle number.
   Again, parameters have been chosen such that the system is in the disordered phase.
   For large densities we find a run time that is approximately proportional to the global density as discussed in Sec.~\ref{sec:implementation}.
   Finding the neighbors for each particle is the part that is proportional to the density.
   Additionally to the determination of neighbors, other operations (that take a run time proportional to the particle number) such as obtaining pseudo random numbers or creating cell lists have to be done.
   Apparently those operations are dominant at small densities. 

   \section{Remarks on versions 1.0 to version 1.2\label{sec:versions}}
   In versions 1.0 and 1.1 there was a mistake in the implementation of reflecting boundary conditions for models with overdamped Langevin dynamics (\textit{additiveL, nonadditiveL, mfL}).
   In that case the orientations of particles have not been reflected when they leave the simulation domain.
   This mistake was not present for Vicsek type models (\textit{VM, NVM, mfVM, mfNVM}) and not for periodic boundary conditions for any model.
   The mistake was corrected with version 1.2.

   Versions 1.0 and 1.1 differ only in the measurements of the moments of the number of neighbors.

   In version 1.0 those moments are calculated from the microscopic number of neighbors $n_j:=|\Omega_j|$.
   If the number of bins for the histogram of the number of neighbors is chosen large enough such that no event is missed, the moments of the histogram are identical to the moments measured directly.
   The number of neighbors of particle $j$, $n_j$, is a microscopic quantity.

   In version 1.1 (and later) the measurement of the moments of the number of neighbors was changed while its histogram is still measured in the same way.
   The behavior described in Subsec.~\ref{subsec:measurements} refers to version 1.1.
   In version 1.1, in each measurement step, the mean number of neighbors $\bar{n}:=\frac{1}{N}\sum_{j=1}^N n_j$ is calculated.
   The first four moments of this quantity are averaged over all measurement time steps.
   Note that now $\bar{n}$ is a macroscopic observable rather than a microscopic quantity.
   The first moment is the same but moments two to four differ (only for the number of neighbors) from version 1.0 to version 1.1.

   It is possible to use version 1.2 to open and analyze a data file produced with versions 1.0 or 1.1.
   All data can be accessed correctly, however a warning is raised, that a different version is used.

   All numerical results presented in Sec.~\ref{sec:examples} have been obtained with version 1.0.
   However, the number of neighbor statistics was not considered in Sec.~\ref{sec:examples} any way and reflecting boundary conditions have been used only for Vicsek type models, such that version 1.2 is supposed to produce (even microscopically) identical results.

   \section{Summary\label{sec:summary}}
   We present a simulation package for the purpose of performing molecular-dynamics-like agent-based simulations of models of aligning self-propelled particles in two dimensions.
   The nature of such prototype models of active matter is different from standard molecular dynamics simulations because the dynamics is not Hamiltonian and $N$-particle interactions of a special type are involved.
   The simulation package is able to simulate models that have been studied by dozens to hundreds of papers, see e.g. \cite{BGHP20, Chate20}, as well as models that have not yet been studied.
   To the best of the authors knowledge, there is no simulation package openly available that can do any of those tasks.
   This paper comes with a few test simulation results of the program for six different models that have been studied before.
   The results of the program are consistent with previous results as well as with mean field theory.

   \section{Acknowledgment}
   The author acknowledges funding through a 'María Zambrano' postdoctoral grant at University of Barcelona financed by the Spanish Ministerio de Universidades and the European Union (Next Generation EU/PRTR).
   The author thanks Universitätsrechenzentrum Greifswald for supporting this work by providing computational resources.

\bibliographystyle{elsarticle-num}

\end{document}